\documentclass{article}
\usepackage{amssymb}
\usepackage{amsmath}
\usepackage{latexsym}
\usepackage{array}
\usepackage{bm}
\usepackage{theorem}
\usepackage{graphicx}
\usepackage{exscale}

\theoremstyle{break}
\newtheorem{Theorem}{Theorem}[section]

\def\Proof{\hfil\break{\bf Proof}\;\;\;\;}
\newtheorem{Proposition}[Theorem]{Proposition}
\newtheorem{Lemma}[Theorem]{Lemma}
\newtheorem{Note}[Theorem]{Note}
\def\DIS{\displaystyle}

\def\n{\boldsymbol{n}}
\def\e{\boldsymbol{e}}
\def\x{\boldsymbol{x}}
\def\y{\boldsymbol{y}}

\def\Z{\mathbb{Z}}

\def\qed{\hfill\hbox{\rule[-2pt]{3pt}{6pt}}}

\begin{document}
\title{Coprimeness-preserving non-integrable extension to the two-dimensional discrete Toda lattice equation}
\author{Ryo Kamiya$^1$, Masataka Kanki$^2$, Takafumi Mase$^1$, Tetsuji Tokihiro$^1$\\
\small $^1$ Graduate School of Mathematical Sciences,\\
\small University of Tokyo, 3-8-1 Komaba, Tokyo 153-8914, Japan\\
\small $^2$ Faculty of Engineering Science,\\
\small Kansai University, 3-3-35 Yamate, Osaka 564-8680, Japan}

\date{}

\maketitle

\begin{abstract}
We introduce a so-called {\em coprimeness-preserving} non-integrable extension to the two-dimensional Toda lattice equation.
We believe that this equation is the first example of such discrete equation defined over a three-dimensional lattice. 
We prove that all the iterates of the equation are irreducible Laurent polynomials of the initial data and that every pair of two iterates is co-prime, which indicate confined singularities of the equation.
By reducing the equation to two- or one-dimensional lattices, we obtain coprimeness-preserving non-integrable extensions to the one-dimensional Toda lattice equation and the Somos-$4$ recurrence.
\end{abstract}
\section{Introduction}
The Toda lattice equation was derived by Toda as a model of one-dimensional chain of masses connected by springs with nonlinear interaction force \cite{Toda}. The Toda lattice is one of the completely integrable systems with multi-soliton solutions.
The equation of motion of the Toda lattice is
\[
\frac{d^2}{dx^2} w_n=\exp (w_{n-1}-w_{n})-\exp (w_n-w_{n+1}),
\]
where $w_n$ is the position of the $n$-th particle.
The time-discretization of the Toda lattice was obtained by Hirota \cite{Hirota} as a bilinear form:
\begin{equation}
\tau_{t+1,n}\tau_{t-1,n}=\tau_{t,n}^2+\tau_{t,n-1}\tau_{t,n+1}. \label{1dtodatau}
\end{equation}
The two-dimensional Toda lattice (2D-Toda) equation was introduced by Mikhailov \cite{Mikhailov}, and Fordy and Gibbons \cite{FG}
and later the elliptic operator case was introduced by Leznov and Saveliev \cite{LS}:
\[
\left(\frac{\partial}{\partial x^2}+\frac{\partial}{\partial y^2}\right)w_n=\exp (w_{n-1}-w_n)-\exp(w_n-w_{n+1}).
\]
Its time-discretization (2D-dToda) was found by Hirota, Tsujimoto and Imai \cite{HTI} as
\begin{equation}
\tau_{t+1,n,m+1}\tau_{t-1,n+1,m}=\tau_{t,n+1,m}\tau_{t,n,m+1}+\tau_{t,n,m}\tau_{t,n+1,m+1}, \label{2dtodabilin}
\end{equation}
where the dependent variable $\tau$ is defined on the three-dimensional lattice $(t,n,m)\in\mathbb{Z}^3$.

There are several integrability criteria of discrete equations. Particularly important two of them are `the singularity confinement test' \cite{SC}, and `the algebraic entropy test' \cite{BV}.
A singularity of a discrete equation is said to be confined, if it is eliminated after a  
finite number of iteration steps, and at that stage, the dependence on the initial data is recovered. 
The discrete equation passes the singularity confinement test (SC test), if all the singularities of the equation are confined.
The SC test was extremely effective in distinguishing `integrable' discrete systems, and also in constructing non-autonomous integrable equations.
For example, discrete analogs of the Painlev\'{e} equations were discovered by searching for non-autonomous extensions to the QRT mappings
that conserve their singularity patterns \cite{RGH}.
However, a `counter example' to SC test has been proposed \cite{HV}.
The equation they have proposed, which is now called the Hietarinta-Viallet equation, passes the SC test, although it is considered to be non-integrable in the sense that it has chaotic orbits of iterates and has no conserved quantities.
They have proposed to use the algebraic entropy to deal with this type of equations.
Algebraic entropy is a quantity to measure the degree growth of the iterates of the equations.
The criterion is that the equation is integrable if and only if its algebraic entropy is zero. This criterion is quite strong: the Hietarinta-Viallet equation has positive ($\log((3+\sqrt{5})/2)>0$) algebraic entropy.
Recently, with the aim of refining these integrability criteria, the authors and others have proposed the `irreducibility' and the `co-primeness' properties to distinguish integrable mappings \cite{dKdVSC2}.
Let us study a discrete mapping whose iterates are always Laurent polynomials of the initial variables. If the equation has this property,
it is said to have the Laurent property \cite{FZ}.
The mapping has the irreducibility, if every iterate is an irreducible Laurent polynomial of the initial variables. Here we assume that the equation is well-defined under a suitable boundary condition.
The equation satisfies the co-primeness condition, if every pair of two iterates is co-prime as Laurent polynomials.
We can also define the co-primeness condition, even if the iterates are not necessarily Laurent polynomials. Moreover, it is also possible to relax the co-primeness condition as follows: the equation passes the co-primeness condition if every pair of two iterates `which is separated by a fixed finite distance' is always co-prime.
The irreducibility and co-primeness are found out to be useful in formulating the integrability of discrete equations defined over the lattice of dimension more than one.
For example, we have proved that these two properties can also be formulated for the discrete Toda equation (both the $\tau$-function form and the nonlinear form) with various boundary conditions: i.e., open, the Dirichlet, and the periodic boundaries \cite{dToda}.

A discrete equation is called `coprimeness-preserving non-integrable', if it passes the singularity confinement test, and at the same time, has exponential growth of the degrees of the iterates.
Note that, in one-dimensional systems, the latter statement is equivalent to saying that the algebraic entropy of the equation is positive, however, for higher-dimensional case we cannot define the entropy in its usual sense.
To construct coprimeness-preserving non-integrable extensions to known discrete equations, we introduce parameters on the powers of the terms.
One of the coprimeness-preserving non-integrable extensions to the discrete KdV equation has been introduced by some of the authors \cite{quasidKdV}, in which these kinds of equations were called `quasi-integrable', however we use another terminology
`coprimeness-preserving non-integrable' in this manuscript.
Some of the equation we study are already introduced by Fomin and Zelevinsky as examples of equations whose iterates are always Laurent polynomials \cite{FZ}. In this manuscript, in addition to the Laurent property, we prove the irreducibility and the co-primeness properties for these equations.
Before starting, let us prepare a small lemma on Laurent polynomials:
\begin{Lemma}\label{lem2}
Let $R$ be a ring of Laurent polynomials. For two Laurent polynomials $f,g\in R$, let us suppose that
$f$ is irreducible in $R$ and, at the same time, $f$ has a non-unit factor which contains a variable that is not in $g$.
Then $f$ and $g$ are co-prime in $R$.
\end{Lemma}
$\because$ ) \ 
Since $f$ is irreducible, the common factor of $f$ and $g$ should be either a unit or $f$ itself.
However, from the assumption, there exists a variable that is in $f$ but not in $g$, thus $f$ cannot be a factor of $g$.
Thus $f$ and $g$ are co-prime.
\qed
\begin{Note}
For example, in a ring $R=\mathbb{Z}[x^{\pm},y^{\pm}]$, $1/xy$ and $1/x^2$ are units because they are invertible in $R$, however, $2\in R$ is not a unit since $1/2\not\in R$.
Another example is that $(x^2+1)/x^3\in R$ is irreducible and is co-prime with $(x^2+3)/x\in R$, but is not co-prime with $(x^4-1)\in R$.
From Lemma \ref{lem2}, $x+y\in R$ is co-prime with $x^2+1\in R$.
\end{Note}
\section{Coprimeness-preserving 2D-dToda}

Our coprimeness-preserving non-integrable extension to the 2D-dToda is given in $\tau$-function form as follows:
\begin{equation}
\tau_{t+1,n,m+1}\tau_{t-1,n+1,m}=\tau_{t,n+1,m}^{k_1}\tau_{t,n,m+1}^{k_2}+\tau_{t,n,m}^{l_1}\tau_{t,n+1,m+1}^{l_2},
\label{pDToda_polinear_eq}
\end{equation}
where $k_1,k_2,l_1,l_2$ are positive integers, with $(k_1,k_2,l_1,l_2)\neq (1,1,1,1)$.
In this manuscript, the set of initial values of equation \eqref{pDToda_polinear_eq} is
$\{\tau_{0,n,m},\tau_{1,n,m}|n,m\in\mathbb{Z}\}$: i.e., set of all the entries in the $t=0$ and $t=1$ planes.
The evolution of \eqref{pDToda_polinear_eq} goes upward in the $t$-axis to $t\ge 2$.
The equation \eqref{pDToda_polinear_eq} is coprimeness-preserving non-integrable in the sense that its degree growth is exponential (as we shall explain in Proposition \ref{2dtodagrowth}) and that it has co-primeness property (Theorem \ref{2dthm}).
\begin{Proposition} \label{2dtodagrowth}
The degrees $\deg(\tau_{t,n,m})$ of the iterates of \eqref{pDToda_polinear_eq} grow exponentially with respect to $t$.
\end{Proposition}
\textbf{Proof}\;\; Let us suppose that the initial values are $\tau_{0,n,m}=1$, $\tau_{1,n,m}=a$ for every $n,m$, and prove that $\deg (\tau_{t,n,m}):=d_t$ grows exponentially.
Since $\tau_t\equiv\tau_{t,n,m}(\forall n,m)$ becomes a polynomial in $a$, and $\tau_{t-1}$ is a factor of $\tau_t$, it is readily obtained that $d_{t+1}=M d_t-d_{t-1}$, with $d_0=0$, $d_1=1$, where $M:=\max(k_1+k_2,l_1+l_2)$. Unless $k_1=k_2=l_1=l_2=1$, we have
\[
\lim_{t\to+\infty} (d_t)^{(1/t)}=\frac{M+\sqrt{M^2-4}}{2}>1.
\]
We have proved an exponential growth for one particular degenerate case, which constitutes the lower bound of the degrees of the iterates.
\qed

Our goal in this paper is to prove the Laurent property, the irreducibility and the co-primeness property of \eqref{pDToda_polinear_eq}, all of which are indication of integrable-like nature of the equation in terms of the singularity analysis.
For the simplicity of our arguments, we will assume that
the greatest common divisor of $(k_1,k_2,l_1,l_2)$ is $1$ or a positive power of $2$,
which is equivalent to the statement that the polynomial
\begin{equation}
x^{k_1}y^{k_2}+z^{l_1}w^{l_2}
\label{irreducibility}
\end{equation}
is irreducible in $\mathbb{Z}[x,y,z,w]$ (This condition shall be elaborated in the appendix, Proposition~\ref{reducibleprop}).
Our main theorem is Theorem \ref{2dthm}, which states that every iterate $\tau_{t,n,m}$ of \eqref{pDToda_polinear_eq}
is an irreducible Laurent polynomial of the initial variables in $\mathbb{Z}$ coefficients, and that two iterates are always co-prime.

We note that the equation \eqref{2dtodabilin} of 2D-dToda is essentially the same as the Hirota-Miwa equation \cite{HirotaMiwa}.
Therefore, we can think of \eqref{pDToda_polinear_eq} as a coprimeness-preserving non-integrable extension to the Hirota-Miwa equation.

\section{Coprimeness-preserving Somos-$4$}
Before dealing with our main target \eqref{pDToda_polinear_eq}, let us
study the properties of a one-dimensional recurrence relation \eqref{Pseudo-SOMOS4}, which is obtained by a reduction of the equation \eqref{pDToda_polinear_eq} on a line: we identify all the iterates $\tau_{t,n,m}$
such that $N=1+2t+n+m$ and introduce a new variable $x_N:=\tau_{t,n,m}$.
The form of the reduced equation is
\begin{equation}\label{Pseudo-SOMOS4}
x_{n+4}x_n=x_{n+3}^lx_{n+1}^m+x_{n+2}^k,
\end{equation}
where $l,m,k$ are positive integers, which we shall call the coprimeness-preserving non-integrable Somos-$4$ sequence.
In fact, \eqref{Pseudo-SOMOS4} is coprimeness-preserving non-integrable for every set of positive integers $l,m,k$ except for $(l,m,k)=(1,1,1)$, $(1,1,2)$.
The flow of the discussion here for the extended Somos-$4$ case helps us to construct the proof of our main theorem.
It is worth noting that equation \eqref{Pseudo-SOMOS4} is given by Fomin and Zelevinsky as the `generalized Somos-$4$ sequence' in their paper (Refer to Example 3.3 in \cite{FZ}. The original Somos-$4$ is the case of $(l,m,k)=(1,1,2)$.), and that the Laurent property of the equation is proved using the ideas in cluster algebras.
Fordy and Marsh obtained \eqref{Pseudo-SOMOS4} in the case of $l=m$ from cluster mutations applied to some cluster mutation-periodic quivers with period one \cite{FordyMarsh}, and Fordy and Hone studied the integrability of such equations including \eqref{Pseudo-SOMOS4} with $l=m$ \cite{FordyHone}. 
Our Proposition \ref{prop1} studies not only the Laurent property but also the irreducibility and co-primeness.
\begin{Proposition}\label{prop1}
Let us assume that the greatest common divisor of $(l,m,k)$ is $1$ or a positive power of $2$.
Then every iterate $x_n$ of equation \eqref{Pseudo-SOMOS4} is an irreducible Laurent polynomial of the initial variables $x_0,x_1,x_2,x_3$. If $n \ne m$, $x_n$ and $x_m$ are co-prime.
\end{Proposition}
We note that the condition on $(l,m,k)$ is equivalent to the irreducibility of the polynomial $x^ly^m+z^k\in\mathbb{Z}[x,y,z]$ (Proposition~\ref{reducibleprop}).
The following Lemma \ref{coprime_lema1} is used to prove Proposition \ref{prop1}.
\begin{Lemma}\label{coprime_lema1}
For every $n\ge 4$ we have
\[
x_n\in R_0:=\mathbb{Z}[x_0^{\pm},x_1^{\pm},x_2^{\pm},x_3^{\pm}],
\]
and that every pair from $x_n,x_{n-1},x_{n-2},x_{n-3}$ is co-prime.
\end{Lemma}
\begin{Note}
The assumption that `$x^ly^m+z^k\in\mathbb{Z}[x,y,z]$ is irreducible' is not used in Lemma \ref{coprime_lema1}.
Thus the Laurent property holds for every positive integer $l,m,k$.
\end{Note}
\Proof
Proof of Lemma \ref{coprime_lema1} is done by induction.
It is easy for $n \le 7$. We just give a proof for $n=7$, assuming Lemma \ref{coprime_lema1} for $n\le 6$. Since $x_7=(x_6^lx_4^m+x_5^k)/x_3$, $x_7$ is trivially a Laurent polynomial of $x_i (i=0,1,2,3)$.
We also obtain the co-primeness of $x_7$ and $x_6$ as follows: if $x_7$ has a common factor (which is not a unit) with $x_6$, that factor must also divide $x_5$, which leads us to a contradiction with the induction hypothesis that $x_6$ is co-prime with $x_5$.

Next we prove the case of larger $n$.
Let us suppose that Lemma \ref{coprime_lema1} is satisfied for every integer less than $n+1$ and prove it for $n+1$.
By continued iterations we have
\begin{align}
&x_{n+1}x_{n-3}=x_{n}^lx_{n-2}^m+x_{n-1}^k \notag\\
&=\left( \frac{x_{n-1}^lx_{n-3}^m+x_{n-2}^k}{x_{n-4}}  \right)^l \left( \frac{x_{n-3}^lx_{n-5}^m+x_{n-4}^k}{x_{n-6}}  \right)^m
+ \left(  \frac{x_{n-2}^lx_{n-4}^m+x_{n-3}^k}{x_{n-5}}  \right)^k \notag\\
&=\frac{x_{n-2}^{kl}x_{n-4}^{km}(x_{n-4}^l x_{n-6}^m+x_{n-5}^k)+O(x_{n-3})}{x_{n-4}^l x_{n-5}^kx_{n-6}^m} \notag\\
&=\frac{x_{n-2}^{kl}x_{n-4}^{km}x_{n-3}x_{n-7}+O(x_{n-3})}{x_{n-4}^l x_{n-5}^kx_{n-6}^m}. \label{lema1calc}
\end{align}
From induction hypotheses, the right hand side of \eqref{lema1calc} ($x_n^l x_{n-2}^m+x_{n-1}^k$) must be a Laurent polynomial of $x_i (i=0,1,2,3)$. 
In equation \eqref{lema1calc}, the term $x_{n-3}$ must be co-prime with $x_{n-4},x_{n-5}$ and $x_{n-6}$.
Thus, by dividing the both sides of \eqref{lema1calc} by $x_{n-3}$, we obtain that the numerator of \eqref{lema1calc}
must be divisible by $x_{n-4}^l x_{n-5}^kx_{n-6}^m$.
Therefore we obtain that $x_{n+1}$ is also a polynomial in $x_i^{\pm}$ ($i=0,1,2,3$).
Since $x_{n+1}x_{n-3}=x_{n}^lx_{n-2}^m+x_{n-1}^k$, $x_{n+1}$ is co-prime with $x_n,x_{n-1},x_{n-2}$.
\qed
Next we prove the irreducibility.
We prepare the following Lemma \ref{lem_put1} to assist the proof for the $n\ge 9$ case.
Let us define $y_n$ as a value of $x_n$ when we substitute
$x_0=x_1=x_2=x_3=1$:
\[
y_n=x_n|_{\{x_0=x_1=x_2=x_3=1\}}.
\]
\begin{Lemma}\label{lem_put1}
The integer sequence $\{y_n\}$ is strictly increasing for $n\geq 3$.
If $l=m=k=1$, we have
\[
y_{10}>y_8y_4>y_9>y_7y_4,
\]
and if otherwise, we have
\[
y_{9} >y_8y_4.
\]
\end{Lemma}
$\because$)
If $k=l=m=1$, we have
\[
y_4=2,\quad y_5=3,\quad y_6=5,\quad y_7=13,\quad y_8=22,\quad y_9=41,\quad y_{10}=111,
\]
and the lemma is readily obtained.
In other cases, we have
\[
y_9-y_8y_4=y_9-2y_8=\frac{y_8^ly_6^m+y_7^k}{y_5}-2y_8=\frac{y_8(y_8^{l-1}y_6^m-2y_5)+y_7^k}{y_5}.
\]
If $l\ge 2$, the right hand side is positive.
In the case of $l=1$, we have $y_6^m=(3+2^k)^m>2y_5$, thus the right hand side is also positive.
\qed

\Proof (\textbf{Proposition} \ref{prop1}) \\
It is sufficient to prove that $x_n$ $(n\ge 4)$ is irreducible.
$x_4=(x_3^l x_1^m +x_2^k)/x_0$ is trivially irreducible because of the assumption.
We use Lemma \ref{lem2previous} on the factorization of Laurent polynomials under a variable transformation, which has been introduced in our previous paper \cite{dKdVSC2}. Lemma \ref{lem2previous} is reproduced in the appendix of this paper. We take
\[
M=4,\ \{q_1,q_2,q_3,q_4\}=\{x_0,x_1,x_2,x_3\},\ \{p_1,p_2,p_3,p_4\}=\{x_1,x_2,x_3,x_4\},
\]
\[
f(x_1,x_2,x_3,x_4)=x_n\ (n\ge 5).
\]
First $x_4$ is trivially a Laurent polynomial of $\{q_1,q_2,q_3,q_4\}$, since $x_4\in R_0$. Also, since equation \eqref{Pseudo-SOMOS4} is invertible,  $x_0$ is a Laurent polynomial of $\{x_1,x_2,x_3,x_4\}$ and is irreducible.
Thus we can factorize $x_n\ (n\ge 5)$ as
\[
x_n=x_4^\alpha f_{irr},
\]
where $f_{irr}$ is some irreducible Laurent polynomial in $R_0$.
Since we have already proved that $x_m\in R_0$, the parameter $\alpha$ must be a non-negative integer.
From Lemma \ref{coprime_lema1}, $x_4$ is co-prime with $x_5, x_6, x_7$, thus we have $\alpha=0$．
Therefore $x_5,x_6,x_7$ are irreducible.
For $x_n\ (n\ge 8)$, Lemma \ref{coprime_lema1} does not tell us if $x_8$ is co-prime with $x_4$, thus we take another approach.
We will prove that, if suitable initial values are taken for $(x_0,x_1,x_2,x_3)\in\mathbb{C}^4$, then we have at the same time $x_4=0$ and $x_8\neq 0$.
Then we can conclude that $x_8$ does not contain a factor $x_4$ when factorized, and thus $\alpha=0$.
Let us investigate the case of $x_8$.
By a direct computation, we have
\begin{align}
x_8&=\frac{1}{x_1^mx_2^kx_3^l}\left\{x_3^{km}x_5^{kl}x_0+lx_2^kx_3^{km}x_5^{k(l-1)}x_6^lx_4^{m-1}+\right. \notag \\
&\qquad + \left. mx_2^{k+m}x_3^{k(m-1)}x_5^{kl}x_4^{l-1}+kx_1^mx_3^{(k-1)m+l}x_5^{(k-1)l}x_4^{k-1}+O(x_4)\right\}. \label{x_8_express}
\end{align}
Since $\DIS x_4=\frac{x_3^lx_1^m+x_2^k}{x_0}$, the value of $x_4$ is zero, when we take $x_0=x_1=x_3=1,\,x_2=t$, $t=\e^{\sqrt{-1}\pi/k}$ as initial values. In this case,
\[
x_5 = \frac{x_3^k}{x_1}=1,\quad x_6 = t^{-1},
\quad x_7 = \frac{x_5^k}{x_3}=1,
\]
and from \eqref{x_8_express},
\begin{equation}\label{eq_x8_1}
x_8=(-1)\left\{ 1+\delta_{m,1}(-1)lt^{-l}+\delta_{l,1}(-1)mt^m+ k \delta_{k,1} \right\},
\end{equation}
where $\delta_{p,q}$ is the Kronecker delta.
From \eqref{eq_x8_1}, we have $x_8\neq 0$ for every $(k,l,m)\in\mathbb{Z}_{>0}^3$ with the exception of $(k,l,m)=(1,1,2),(1,2,1),(3,1,1)$.
We can study these three cases separately and can find at least one set of $(x_0,x_1,x_2,x_3)\in\mathbb{C}^4$ such that $x_4=0$ and $x_8\neq 0$.
(In the case of $(k,l,m)=(3,1,1)$, for example,
if we take the initial values as $x_1=-1,x_0=x_2=x_3=1$,
we have $x_4=0$ and $x_8=3\neq 0$.)
Thus the irreducibility of $x_8$ is proved. 

The iterate $x_n (n\ge 9)$ has the following two factorizations from Lemma \ref{lem2previous},
\begin{equation}
x_n=x_4^\alpha f_{irr}=x_5^{\beta_1}x_6^{\beta_2}x_7^{\beta_3}x_8^{\beta_4} g_{irr}\ \ (n\ge 9), \label{xnx4}
\end{equation}
where $f_{irr},\,g_{irr}$ are both irreducible Laurent polynomials of the initial variables.
To obtain \eqref{xnx4}, we have chosen $\{p_1,p_2,p_3,p_4\}=\{x_5,x_6,x_7,x_8\}$ for the second equality and have applied Lemma \ref{lem2previous}.
Let us suppose that $x_n (n\ge 9)$ is reducible and derive a contradiction.
From \eqref{xnx4}, a factorization of $x_n$ is limited to the following type:
\[
x_n=x_4 x_j \times \mbox{unit}  \qquad (n\ge 9,\, j \in \{5,6,7,8\}),
\]
where `unit' is a unit element in $R_0$.
When we substitute $x_0=x_1=x_2=x_3=1$ in the above equation, the `unit' goes to $1$ and we have
\[
y_n=y_4y_j \qquad (n\ge 9,\, j \in \{5,6,7,8\}),
\]
which is impossible from Lemma \ref{lem_put1}.
Thus $x_n (n\ge 9)$ is irreducible.
We have completed the proof that $x_n$ is irreducible for every $n\ge 1$.
Since the sequence $\{y_n\}$ is strictly increasing for $n\geq 3$, two iterates $x_n$ and $x_{n'}$ with $n\neq n'$  cannot be equal to each other.
Two irreducible distinct elements must be co-prime, and the proof of Proposition \ref{prop1} is finished.
\qed
\section{Co-primeness of extended 2D-dToda}
Next we move on to our main equation \eqref{pDToda_polinear_eq}.
For ease of notation, let us shift all the variables $\tau_{t,n,m}$ to
$\tau_{t,n+t/2,m-t/2}$.
These shifts produce half-integer lattice points, however, the evolution of equation \eqref{pDToda_polinear_eq} is simplified since it is now described using six vertices of an octahedron.
For simplicity let us define the following symbols in $(n,m)$-plane:
$\n=(n,m)$ and
\[
\e_1=\left(\frac{1}{2},\frac{1}{2}\right),\ \e_2 = \left(-\frac{1}{2},\frac{1}{2}\right),\ \e_3=\left(-\frac{1}{2},-\frac{1}{2}\right),\ \e_4 = \left(\frac{1}{2},-\frac{1}{2}\right).
\]
From here on, $\hat{\tau}$ denotes a up-shift in the $t$-axis, and $\check{\tau}$ denotes a downshift in the $t$-axis.
Then equation \eqref{pDToda_polinear_eq} can be expressed as
\begin{equation}\label{new_Toda_eq}
\hat{\tau}_{\n} \check{\tau}_{\n}=\tau_{\n+\e_4}^{k_1}\tau_{\n+\e_2}^{k_2}+\tau_{\n+\e_3}^{l_1}\tau_{\n+\e_1}^{l_2}.
\end{equation}
We have the following main theorem on the irreducibility and co-primeness of 2D-dToda:
\begin{Theorem} \label{2dthm}
Let us assume that the greatest common divisor of $(k_1,k_2,l_1,l_2)$ is a non-negative power of $2$. Then each iterate $\tau_{t,\n}$ of equation \eqref{new_Toda_eq} is an irreducible Laurent polynomial of the initial variables
\[
\left\{\tau_{t=0, \n},\,\tau_{t=1, \boldsymbol{m}}\, \Big|\, \n\in\mathbb{Z}^2, \boldsymbol{m}\in\left(\mathbb{Z}+\frac{1}{2}\right)^2\right\}.
\]
Every pair of the iterates is always co-prime.
\end{Theorem}

\Proof 
We will rewrite $x_{\n}:=\tau_{t=0, \n}$, $y_{\n}:=\tau_{t=1, \n}$, and so on: i.e., we use
$z_{\n},u_{\n},v_{\n},w_{\n},p_{\n},q_{\n}$ for the values of $\tau_{t=i, \n}$ at $i=2,3,4,5,6,7$.
We will prove Theorem \ref{2dthm} step by step from $x_{\n}$ to $p_{\n}$ and beyond.
\begin{enumerate}
\item The case of $t=2$:
If $\n \ne \n'$, then $z_{\n}$ and $z_{\n'}$ are irreducible in
\[
R:=\Z\left[\x_{\n}^{\pm}, \y_{\n'}^{\pm}\, \Big|\, \n\in\mathbb{Z}^2, \n'\in\left(\mathbb{Z}+1/2\right)^2\right],
\]
and are co-prime.
\par
$\because$ )
Since
\[
z_{\n}=\frac{1}{x_{\n}}\left(y_{\n+\e_4}^{k_1}y_{\n+\e_2}^{k_2}+y_{\n+\e_3}^{l_1}y_{\n+\e_1}^{l_2}\right),
\]
if $\n \ne \n'$, two iterates $z_{\n}$ and $z_{\n'}$ have at most two variables in the $(t=1)$-plane ($y_*$) in common.
Thus from Lemma \ref{lem2}, two iterates must be co-prime.
\qed
\item The case of $t=3$:
if $\n \ne \n'$, two iterates $u_{\n}$ and $u_{\n'}$ are irreducible in $R=\Z[\x^{\pm}, \y^{\pm}] $ and are co-prime.
Each $u_{\n}$ is co-prime with $z_{\n'}$ for all $\n'\in\mathbb{Z}^2$.
\par
$\because$ )
From
\[
u_{\n}=\frac{1}{y_{\n}}\left(z_{\n+\e_4}^{k_1}z_{\n+\e_2}^{k_2}+z_{\n+\e_3}^{l_1}z_{\n+\e_1}^{l_2}\right)
\]
and from Lemma \ref{lem2previous}, we obtain the factorization of $u_{\n}$ as
\[
u_{\n}=\left( \prod_{k\, \mbox{finite}} z_{\n_k}^{\alpha_k}\right) f_{irr}\qquad (\n_k\in \mathbb{Z}^2,\, \alpha_k \in \Z_{\ge 0}),
\]
where $f_{irr}$ is irreducible in $R$.
Since $u_{\n}$ does not have the factor $z_{\n+\e_i}$ for $i=1,2,3,4$,
we have $\alpha_k=0$ if $\n_k = \n+\e_i$ ($i=1,2,3,4$).
(Note that $\{z_{\n}\}$ is mutually co-prime, and thus, two distinct
$z_*$'s are not identical.)
For other $\n_k$, the iterate
$z_{\n_k}$ contains at least one term $y_{\n_k+\e_i}$ that does not appear in $u_{\n}$.
Since $z_{\n_k}$ is binomial with respect to the variables $\{y_{\n}\}$ in $t=1$, the term $y_{\n_k+\e_i}$ cannot be eliminated by multiplying some unit element in $R$.
Thus from Lemma \ref{lem2}, two iterates $z_{\n_k}$ and $u_{\n}$ are co-prime, and we have $\alpha_k=0$. We have proved that $u_{\n}$ is irreducible.
It is readily obtained that each $u_{\n}$ is co-prime with $z_{\n'}$ for every $\n'$.
The final step is to prove that $u_{\n}$ and $u_{\n'}$ are co-prime if $\n\neq \n'$.
Each iterate $u_{\n}$, when expanded as a Laurent polynomial in $R$, contains nine terms $y_{*}$ in $(t=1)$-plane, none of which is cancelled out
by multiplying unit elements in $R$. When $\n \ne \n'$, there must be at least one term $y_{*}$ that does not appear simultaneously in the iterates $u_{\n}$ and $u_{\n'}$. Therefore, using Lemma \ref{lem2}, we obtain the co-primeness of $u_{\n}$ and $u_{\n'}$.
\qed
\item The case of $t=4$ (Part I):
$v_{\n}\in R$:
\par
$\because$ )
Let us denote 9 points $a,b,c,\ldots,i$ in the lattice plane $\mathbb{Z}^2$ on which the variables $v_{\n}, z_{\n}, x_{\n}$ lie,
and denote 8 points $\alpha,\beta,\gamma,\delta,\alpha',\beta',\gamma',\delta'$ in the lattice plane $(\mathbb{Z}+1/2)^2$ on which the variables $u_{\n}, y_{\n}$ lie,
as in Figure \ref{LatticePoints}. Let us take the point `$e$' at the center as $\n=e$.
\begin{figure}[h]
\begin{center}
 \includegraphics[bb=48 88 306 316,width=8.0cm]{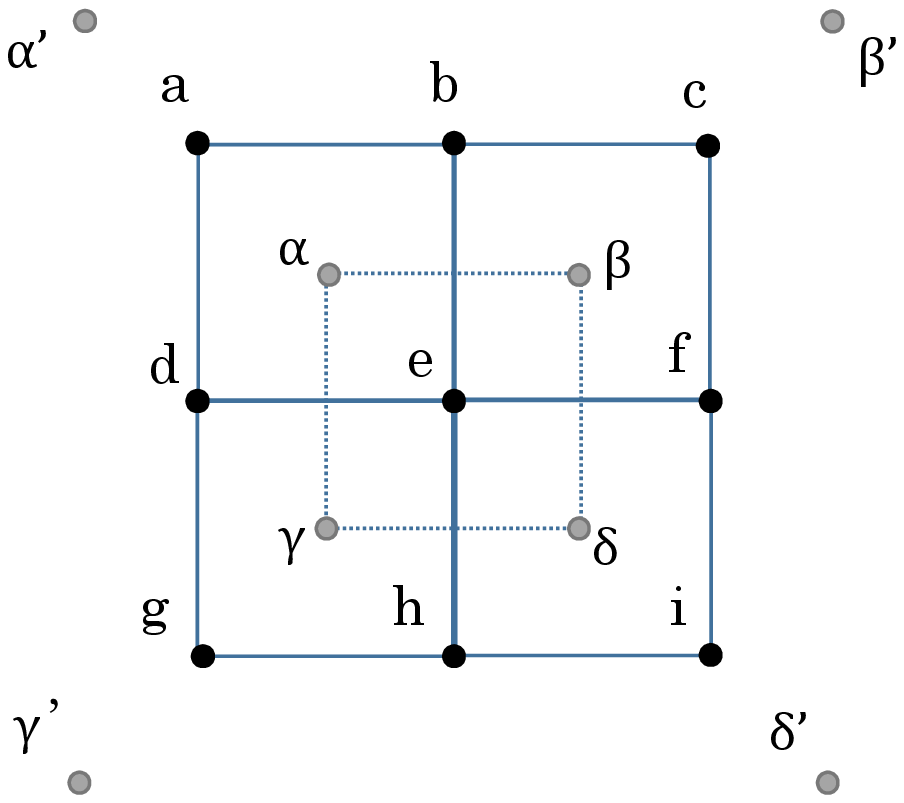}
\caption{Numbering of the lattice points}
\label{LatticePoints}
\end{center}
\end{figure}
By a direct computation, we have
\begin{align}
z_e v_e&=u_\delta^{k_1}u_\alpha^{k_2}+u_\gamma^{l_1}u_{\beta}^{l_2}\notag \\
&=\left( \frac{z_i^{k_1}z_e^{k_2}+z_h^{l_1}z_f^{l_2}}{y_\delta}    \right)^{k_1}
\left( \frac{z_e^{k_1}z_a^{k_2}+z_d^{l_1}z_b^{l_2}}{y_\alpha}    \right)^{k_2}\notag \\
&\qquad +\left( \frac{z_h^{k_1}z_d^{k_2}+z_g^{l_1}z_e^{l_2}}{y_\gamma}    \right)^{l_1}
\left( \frac{z_f^{k_1}z_b^{k_2}+z_e^{l_1}z_c^{l_2}}{y_\beta}    \right)^{l_2}\notag \\
&=\frac{(z_h^{l_1k_1}z_f^{l_2k_1} z_d^{l_1k_2}z_b^{l_2k_2} )(y_\gamma^{l_1}y_{\beta}^{l_2}+y_{\delta}^{k_1}y_{\alpha}^{k_2} )+O(z_e) }{y_\delta^{k_1}y_{\alpha}^{k_2}y_{\gamma}^{l_1}y_{\beta}^{l_2}}\notag \\
&=\frac{z_e x_e\cdot z_h^{l_1k_1}z_f^{l_2k_1} z_d^{l_1k_2}z_b^{l_2k_2} +O(z_e)}{y_\delta^{k_1}y_{\alpha}^{k_2}y_{\gamma}^{l_1}y_{\beta}^{l_2}}. \label{zeve2}
\end{align}
By eliminating $z_e$ from both sides, we conclude that $v_e$ is a Laurent polynomial.

The calculation of $\tau_{t,\n}\tau_{t-2,\n}$ can be done for $t \ge 5$ in the same manner as we have done for $t=4$ in \eqref{zeve2}.
Therefore the Laurent property (not the irreducibility) in Theorem \ref{2dthm}
can be partially proved first: i.e., if we suppose that $\tau_{t,\n}\in R$
and is irreducible in $R$ for every $0 \le t \le t_0$, then we readily conclude that $\tau_{t,\n}\in R$ for $t=t_0+1$. However the irreducibility for $t=t_0+1$ needs more careful treatment.
\item The case of $t=4$ (Part II):
Iterates $v_{\n}$ $(\n\in\mathbb{Z}^2)$ are irreducible Laurent polynomials and are mutually co-prime.
They are also co-prime with every $u_{\n'}$ $(\n'\in(\mathbb{Z}+1/2)^2)$ and $z_{\n'}$ $(\n'\in\mathbb{Z}^2)$.
\par
$\because$ )
Let us take $\n=e$ and use the numbering of the lattice points in Figure \ref{LatticePoints}. We will prove that $v_e$ is an irreducible Laurent polynomial.
By using Lemma \ref{lem2previous}, we obtain the following factorization of $v_e$:
\[
v_{e}=\left(\prod_{\n'}z_{\n'}^{\alpha_{\n'}}\right) f_{irr}\quad(\alpha_{\n'} \in \Z_{\ge 0}),
\]
where $f_{irr}$ is irreducible in $R$.
For $\n'$ with $\n'\notin\{a,b,...,i\}$, there exists a term $y_*$ of $t=2$, that is contained in the iterate $z_{\n'}$, and at the same time, is not contained in $v_e$. This term $y_*$ cannot be cancelled out by multiplying some unit element, and thus from Lemma \ref{lem2} we have $\alpha_{\n'}=0$.

The proof is completed if we prove that $\alpha_{\n'}=0$ for $\n'\in\{a,b,...,i\}$.
Let us define two ideals of $R$ as $I_1:=z_a\cdot R$ and $I_2:=z_e^2\cdot R$.
From the symmetry of the evolution equation and the configuration of the variables, it is sufficient to prove that $\alpha_a=\alpha_b=\alpha_e=0$.

\paragraph{(Proof of $\alpha_a=0$)}
Since $z_e$ and $z_a$ are co-prime, we only have to prove that $z_ev_e\notin I_1$. Let us suppose that $z_ev_e\in I_1$ and lead us to a contradiction.
Note that $z_a$ does not have a term $y_{\delta'}$ when written with the initial variables. Thus it is necessary that the term of the highest order of $z_ev_e$ be divisible by $z_a$, when the terms of $z_e v_e$ is re-arranged with respect to $y_{\delta'}$ (as a Laurent polynomial of $y_{\delta'}$).
We have
\begin{equation}
z_ev_e=u_\delta^{k_1}u_\alpha^{k_2}+u_\gamma^{l_1}u_{\beta}^{l_2}=\left( \frac{z_i^{k_1}z_e^{k_2}+z_h^{l_1}z_f^{l_2}}{y_\delta}    \right)^{k_1}
u_\alpha^{k_2}
+u_\gamma^{l_1}u_{\beta}^{l_2}. \label{zeve1}
\end{equation}
Among the eight terms $z_i,\ z_e,\ z_h,\ z_f$, $y_\delta,\ u_\alpha,\ u_\gamma,\ u_\beta$ that appear in \eqref{zeve1},
only $z_i$ contains the term $y_{\delta'}$ in its expansion. Therefore when we re-arrange the terms of $z_ev_e$ as a Laurent polynomial of the variable $y_{\delta'}$, its degree is $k_1^3$. The coefficient of the term $y_{\delta'}^{k_1^3}$ is equal to
\[
x_i^{-k_1^2}y_\delta^{k_1^2k_2-k_1}z_e^{k_1k_2}u_\alpha^{k_2},
\]
and it should be divisible by $z_a$.
This leads us to a contradiction because we have already proved that $u_\alpha$ and $z_e$ are both co-prime with $z_a$.
We have $z_ev_e\notin I_1$ and thus $\alpha_a=0$.

Proof of $\alpha_b=0$ can be done in a similar manner to the previous step and is omitted in this paper.

\paragraph{(Proof of $\alpha_e=0$)}
Let us suppose that $z_ev_e\in I_2$.
The four variables $y_{\alpha'},y_{\beta'},y_{\gamma'},y_{\delta'}$ are not used to construct $z_e$.
Thus, when an element of $I_2$ is considered as a Laurent polynomial in $y_{\alpha'}$, the coefficient of its highest term should be divisible by $z_e^2$.
By further expanding the iterates $u_\alpha, u_\beta, u_\gamma$ in the right hand side of equation \eqref{zeve1}, we have
\begin{align}
I_2&=z_ev_e+I_2\notag\\
&=y_\delta^{-k_1}y_\alpha^{-k_2}(k_1z_i^{k_1}z_e^{k_2}z_h^{l_1(k_1-1)}z_f^{l_2(k_1-1)}+z_h^{l_1k_1}z_f^{l_2k_1})\notag\\
&\qquad\times(k_2z_e^{k_1}z_a^{k_2}z_d^{l_1(k_2-1)}z_b^{l_2(k_2-1)}+z_d^{l_1k_2}z_b^{l_2k_2})\notag\\
&\qquad+y_\gamma^{-l_1}y_\beta^{-l_2}(l_1z_g^{l_1}z_e^{l_2}z_h^{k_1(l_1-1)}z_d^{k_2(l_1-1)}+z_h^{k_1l_1}z_d^{k_2l_1}) \notag\\
&\qquad\times(l_2z_e^{l_1}z_c^{l_2}z_f^{k_1(l_2-1)}z_b^{k_2(l_2-1)}+z_f^{k_1l_2}z_b^{k_2l_2})+I_2 \notag\\
&=y_\delta^{-k_1}y_\alpha^{-k_2}(k_1z_i^{k_1}z_e^{k_2}z_h^{l_1(k_1-1)}z_f^{l_2(k_1-1)}z_d^{l_1k_2}z_b^{l_2k_2}\notag\\
&\qquad+k_2z_e^{k_1}z_a^{k_2}z_d^{l_1(k_2-1)}z_b^{l_2(k_2-1)}z_h^{l_1k_1}z_f^{l_2k_1})\notag \\
&\qquad+y_\gamma^{-l_1}y_\beta^{-l_2}(l_1z_g^{l_1}z_e^{l_2}z_h^{k_1(l_1-1)}z_d^{k_2(l_1-1)}z_f^{k_1l_2}z_b^{k_2l_2}\notag\\
&\qquad+l_2z_e^{l_1}z_c^{l_2}z_f^{k_1(l_2-1)}z_b^{k_2(l_2-1)}z_h^{k_1l_1}z_d^{k_2l_1})\notag \\
&\qquad+(y_\delta^{-k_1}y_\alpha^{-k_2}+y_\gamma^{-l_1}y_\beta^{-l_2})z_h^{l_1k_1}z_f^{l_2k_1}z_d^{l_1k_2}z_b^{l_2k_2}
+I_2. \label{aezero2}
\end{align}
Here the term with $z_e^2$ is absorbed in the ideal $I_2$.
Among all the iterates on $t=2$ plane (i.e., $z_{\n}$),
the only iterate that contain $y_{\alpha'}$ in its expansion is $z_a$.
Let us re-arrange the right hand side of \eqref{aezero2}
as a Laurent polynomial of $y_{\alpha'}$.
Then the coefficient of the highest order ($k_2^2$-th order) is
\begin{equation*}
k_2x_a^{-k_2}y_\delta^{-k_1}y_\alpha^{k_1k_2-k_2}z_e^{k_1}z_d^{l_1(k_2-1)}z_b^{l_2(k_2-1)}z_h^{l_1k_1}z_f^{l_2k_1},
\end{equation*}
which should be divisible by $z_e^2$.
Since every pair of two terms of $z_{\n}$ is co-prime, we conclude that $k_1\geq 2$.
The same arguments also show that $l_1\geq2,\ l_2\geq 2$ and $k_2\geq2$.
(We consider \eqref{aezero2} as a Laurent polynomial of $y_{\beta'}$, $y_{\gamma'}$, and $y_{\delta'}$ each.)
Therefore, in \eqref{aezero2}, the first two terms are divisible by $z_e^2$ and belong to the ideal $I_2$. Thus the last term
$(y_\delta^{-k_1}y_\alpha^{-k_2}+y_\gamma^{-l_1}y_\beta^{-l_2})z_h^{l_1k_1}z_f^{l_2k_1}z_d^{l_1k_2}z_b^{l_2k_2}$ of \eqref{aezero2} is also in $I_2$.
On the other hand, from the evolution equation, we have
\begin{align}
&(y_\delta^{-k_1}y_\alpha^{-k_2}+y_\gamma^{-l_1}y_\beta^{-l_2})z_h^{l_1k_1}z_f^{l_2k_1}z_d^{l_1k_2}z_b^{l_2k_2}\notag\\
=&y_\delta^{-k_1}y_\alpha^{-k_2}y_\gamma^{-l_1}y_\beta^{-l_2}z_h^{l_1k_1}z_f^{l_2k_1}z_d^{l_1k_2}z_b^{l_2k_2}x_e\cdot z_e, \label{aezero3}
\end{align}
which indicates that \eqref{aezero3} is divisible by $z_e$ only once. This leads us to a contradiction.
Thus $z_ev_e\notin I_2$, and therefore $\alpha_e=0$ is proved.

Now we have finished the proof of the irreducibility of $v_e$.

Next let us prove that each $v_{\n}$ is co-prime with every iterates below $t=4$.
Let us substitute $x_{\n}\to 1,\,y_{\n}\to 1$. Then $\tau_{\n}$ is a constant independent of a choice of $\n$ for a fixed $t$:
 we define $\tilde{\tau}:=\tau_{\n}|_{x_{\n}\to 1,y_{\n}\to 1}$ and use symbols such as $\tilde{z}$ for $\tilde{\tau} (t=2)$ and so on.
 Then we have 
$\tilde{z}=2$, $\tilde{u}=2^{k_1+k_2}+2^{l_1+l_2}$,
$\tilde{v}=(\tilde{u}^{k_1+k_2}+\tilde{u}^{l_1+l_2})/2$,
which indicates that $\tilde{v}> \tilde{u},\,\tilde{z}$.
Therefore $v_{\n}$ cannot have a common factor with $\{u_{\n'}\}$ or $\{z_{\n'}\}$.
Finally we note that $v_{\n}$ and $v_{\n'}$ are co-prime if $\n \ne \n'$. This is readily proved using Lemma \ref{lem2}, since $v_{\n}$ and $v_{\n'}$ are Laurent polynomials of the same degree, and they have distinct terms $y_*$ in the $t=1$ plane.
\qed

\item Proof of the case $t=5$:
Let us prove that $w_{\n}$ is an irreducible Laurent polynomial in $R$ and every pair is co-prime.
Also we prove that $w_{\n}$ is co-prime with $v_{\n'},\,u_{\n'},\,z_{\n'}$.

$\because$) Let us use Lemma \ref{lem2previous} to consider the possible factorizations of $w_{\n}$ as we have done for the coprimeness-preserving non-integrable Somos-$4$ sequence
$(n\ge 9)$ in Proposition \ref{prop1}.
Let us suppose that $w_{\n}$ is reducible.
Then we have only two types of factorizations as follows:
\[
w_{\n}=\mbox{unit}\times z_{\n'} u_{\n''}\; \mbox{or}\; w_{\n}=\mbox{unit}\times z_{\n'} v_{\n''}.
\]
However, since $\tilde{w}>\tilde{z}\tilde{v}>\tilde{z}\tilde{u}$, we have a contradiction. Thus $w_{\n}$ is irreducible.
By a discussion similar to that in the previous part, we conclude that $w_{\n}$ and $w_{\n'}$ are co-prime if $\n \ne \n'$.
\qed
\item The proof of the case $t=6$:
All the iterates $p_{\n}$ are irreducible Laurent polynomials in $R$ and are pairwise co-prime.
Moreover they are co-prime with $w_{\n'},\, v_{\n'},\,u_{\n'},\, z_{\n'}$.
\par
$\because$) The discussion proceeds in the same way as in the previous part.\qed
\item The proof of the case $t \ge 7$: For $t\ge 7$, each term $\tau_{t,\n}$ is an irreducible Laurent polynomial in $R$ and is co-prime with every iterate $\tau_{s,\n'}$ ($s \le t$).
\par
$\because$) From the discussion in the case of $t=4$ (Part I), the iterate $q_{\n}\in R$ for every $t\ge 7$. By using Lemma \ref{lem2previous} for $q_{\n}$, we have three types of factorizations:
\begin{align*}
q_{\n}&=\left(\prod _{\n'}z_{\n'}^{\alpha_{\n'}}\right) f_{irr} \\
&=\left(\prod _{\n'}u_{\n'}^{\beta_{\n'}}\right) \left(\prod_{\n'}v_{\n'}^{\beta_{\n'}'}\right)g_{irr} \\
&= \left(\prod _{\n'}w_{\n'}^{\gamma_{\n'}}\right) \left(\prod_{\n'}p_{\n'}^{\gamma_{\n'}'}\right) h_{irr}, 
\end{align*}
where $f_{irr}$, $g_{irr}$, $h_{irr}$ are irreducible in $R$.
These factorizations cannot be compatible unless $q_{\n}$ is irreducible in $R$ (in that case, $\alpha_{\n'}=\cdots =\gamma'_{\n'}=0$ and $f_{irr}=g_{irr}=h_{irr}$).
By the same argument to the previous step, each pair of $q_{\n}$ and $q_{\n'}$ is co-prime if $\n \ne \n'$.

\item The proof for $t \ge 8$ is done inductively.
\qed
\end{enumerate}

\section{Coprimeness-preserving 1D-dToda}
The following equation \eqref{DToda_polinear_eq} is obtained from a reduction of equation \eqref{pDToda_polinear_eq} to a two-dimensional lattice.
Let us make a transformation $n+m\to N$ and identify all $\tau_{t,n,m}$ with $N=n+m$. Then $\tau_{t,N}:=\tau_{t,n,m}$ satisfy
\begin{equation}
\tau_{t+1,N}\tau_{t-1,N}=\tau_{t,N}^{k}+\tau_{t,N-1}^{l_1}\tau_{t,N+1}^{l_2}\qquad (k, l_1, l_2 \in \Z_+).
\label{DToda_polinear_eq}
\end{equation}
For an arbitrary $(k,l_1,l_2)\in\mathbb{Z}_{+}$, equation \eqref{DToda_polinear_eq} passes the singularity confinement test and has irreducibility and co-primeness properties.
If $(k,l_1,l_2)\neq (1,1,1),(2,1,1)$, the equation \eqref{DToda_polinear_eq}
has exponential growth of the degrees of its iterates.
The equation \eqref{DToda_polinear_eq} is the discrete Toda equation \eqref{1dtodatau} if $k=2$, $l_1=1$, $l_2=1$.
In the case of $(k,l_1,l_2)=(1,1,1)$, we can prove that the degree of its iterates grows according to a polynomial of degree one, by applying a discussion in \cite{Mase}.
We note that equation \eqref{DToda_polinear_eq} is already mentioned in \cite{FZ} as `Number walls' and its Laurent property is proved.
We shall call \eqref{DToda_polinear_eq} a `coprimeness-preserving non-integrable 1D discrete Toda equation' and include it in the category of coprimeness-preserving non-integrable systems.
\begin{Proposition} \label{1dthm}
Let us define the evolution of the equation \eqref{DToda_polinear_eq} from
the initial variables $\tau_{0,n}$, $\tau_{1,n}$ $(n\in\mathbb{Z})$, upward on the $t$-axis.
Then every iterate $\tau_{t,n}$ for $t\ge 3$ is an irreducible Laurent polynomial in
\[
\mathbb{Z}\left[\tau_{0,n}^{\pm},\tau_{1,n}^{\pm}\, |\, n\in \mathbb{Z}\right],
\]
and every pair of the iterates is co-prime.
\end{Proposition}
Note that the proof of Proposition \ref{1dthm} is not directly transferred from that of Theorem \ref{2dthm}, since the irreducibility and the co-primeness are not necessarily conserved under the reduction.
Proof of this proposition is omitted in this paper, since it can be done inductively with respect to $t$, with the help of Lemma \ref{lem2previous}.

\section{Conclusion and Discussion}
In this paper we have constructed a coprimeness-preserving non-integrable extension to the two-dimensional discrete Toda equation (Cp-2D-dToda), and proved that it has Laurent property, the irreducibility and the co-primeness property.
The Cp-2D-dToda equation is considered to be the first example of coprimeness-preserving non-integrable equations defined on a three-dimensional lattice $\mathbb{Z}^3$.
We have also presented the coprimeness-preserving non-integrable Somos-$4$ recurrence (Cp-Somos-$4$), the coprimeness-preserving non-integrable 1D discrete Toda equation (Cp-1D-dToda), through a reduction from Cp-2D-dToda, and have proved that they also have the Laurent property, the irreducibility and the co-primeness properties (although some parts of the proofs have been omitted).
The Cp-Somos-$4$ and Cp-1D-dToda are already known to have the Laurent property \cite{FZ}, and in our paper, we have added the proof for the irreducibility and the co-primeness.

Properties of the irreducibility and co-primeness are considered as strong indications of the discrete integrability (including coprimeness-preserving yet non-integrable cases), and also are algebraic interpretations of singularity confinement.
It is expected that further exploration into the topics related to the co-primeness, will lead us to constructing refined integrability criteria for discrete dynamical systems. For example, in some discrete equations, even when the general iterates are not irreducible, we can still prove that every pair of two iterates are co-prime. This phenomenon might lie in the boundary of integrable systems and non-integrable ones.
For example, even if we remove the condition on $(l,m,k)$, the Cp-Somos-$4$ equation {\bf does} satisfy the co-primeness property, although its iterates are reducible in general.
Another interesting topic to study is the nonlinear forms of the equations that we have investigated here.
The original 2D-dToda equation has the $\tau$-function bilinear form \eqref{2dtodabilin} as we have presented in this article, and also has a nonlinear expression related to it. We hope to formulate a coprimeness-preserving non-integrable extension to the nonlinear 2D-dToda, and prove the co-primeness property for the nonlinear equation.
We will also try to construct extensions to other bilinear equations.
These works are expected to be presented in our subsequent works.
 
\section*{Acknowledgments}
The authors thank Professors A. P. Fordy and R. Willox for useful comments.
This work is partially supported by KAKENHI Grant Numbers 15H06128, 16H06711, JP21340034.

\appendix
\section{Conditions on the indices of the equation \eqref{pDToda_polinear_eq}}
We prove the following proposition:
\begin{Proposition}\label{reducibleprop}
Let $r$ be a positive integer, and let $a_1,a_2,a_3,a_4$ be non-negative integers with GCD$(a_1,a_2,a_3,a_4)=1$.
Then the polynomial
\[
X_1^{a_1 r} X_2^{a_2 r} + X_3^{a_3 r} X_4^{a_4 r}
\]
is irreducible in $\mathbb{Z}[X_1,X_2,X_3,X_4]$ if and only if $r=2^l$ $(l\ge 0)$.
\end{Proposition}
\textbf{Sketch of Proof}\;\;
If $r$ has a odd prime factor, it is trivial that $X_1^{a_1 r} X_2^{a_2 r} + X_3^{a_3 r} X_4^{a_4 r}$ is reducible.
We prove that, if $r$ is a power of $2$, $X_1^{a_1 r} X_2^{a_2 r} + X_3^{a_3 r} X_4^{a_4 r}$ is irreducible.
It is sufficient to prove the irreducibility of
\[
F:=(X_1^{a_1}X_2^{a_2}X_3^{-a_3}X_4^{-a_4})^r+1
\]
as a Laurent polynomial.
From Lemma \ref{lem21}, there exists a matrix $B=(b_{ij})\in \operatorname{GL}_{\mathbb{Z}}(4)$ such that
\[
	b_{11} = a_1, \quad
	b_{21} = a_2, \quad
	b_{31} = -a_3, \quad
	b_{41} = -a_4.
\]
Then the following ring homomorphism
\[
	\psi \colon \mathbb{Z}[Y^{\pm}_1, Y^{\pm}_2, Y^{\pm}_3, Y^{\pm}_4] \to \mathbb{Z}[X^{\pm}_1, X^{\pm}_2, X^{\pm}_3, X^{\pm}_4],
\]
defined by
\[
	Y_i \mapsto X^{b_{i1}}_1 X^{b_{i2}}_2 X^{b_{i3}}_3 X^{b_{i4}}_4,
\]
is in fact an isomorphism, since we can define its inverse by
\[
	\psi^{-1} \colon \mathbb{Z}[X^{\pm}_1, X^{\pm}_2, X^{\pm}_3, X^{\pm}_4] \to \mathbb{Z}[Y^{\pm}_1, Y^{\pm}_2, Y^{\pm}_3, Y^{\pm}_4], \quad
	X_i \mapsto Y^{c_{i1}}_1 Y^{c_{i2}}_2 Y^{c_{i3}}_3 Y^{c_{i4}}_4,
\]
where $C=(c_{ij})=B^{-1}$.
From the definition of the map $\psi$, we have
\[
	\psi(Y^r_1 + 1) = F.
\]
Since $r$ is a non-negative power of $2$, $Y^r_1+1$ is the $2r$-th cyclotomic polynomial and is irreducible. 
The irreducibility is preserved under the isomorphism $\psi$, thus $F$ is irreducible.

\begin{Lemma}\label{lem21}
Suppose that $a_1, \ldots, a_N \in \mathbb{Z}$ are co-prime integers.
Then there exists a matrix $B = (b_{ij}) \in \operatorname{GL}_{\mathbb{Z}}(N)$ such that
\[
	b_{i1} = a_i,
\]
for every $i$.
\end{Lemma}
Proof can be done using a knowledge of elementary algebra.

\section{Lemma on the factorization of Laurent polynomials in \cite{dKdVSC2}}
Let us reproduce a lemma on how the Laurent polynomial is factorized when we make a transformation to the variables, where the two sets of variables (before and after the transformation) satisfy some good conditions. In usual settings, the conditions are satisfied thanks to the Laurent property and the invertibility of the equation.
\begin{Lemma}[\cite{dKdVSC2}]
\label{lem2previous}
Let $M$ be a positive integer and let $\{p_1,p_2,\cdots,p_M\}$ and $\{q_1,q_2,\cdots ,q_M\}$ be two sets of independent variables with the following properties:
\begin{align*}
p_j &\in \mathbb{Z}\left[ q_1^{\pm}, q_2^{\pm},\cdots ,q_M^{\pm}\right], 
q_j \in \mathbb{Z}\left[ p_1^{\pm}, p_2^{\pm},\cdots ,p_M^{\pm}\right], \\
q_j&\ \mbox{is irreducible as an element of}\ \mathbb{Z}\left[ p_1^{\pm}, p_2^{\pm},\cdots ,p_M^{\pm}\right],  \notag
\end{align*}
for $j=1,2,\cdots, M$.
Let us take an irreducible Laurent polynomial
\[
f(p_1,\cdots, p_M)\in \mathbb{Z}\left[ p_1^{\pm}, p_2^{\pm},\cdots ,p_M^{\pm}\right],
\]
and another (not necessarily irreducible) Laurent polynomial
\[
g(q_1,\cdots, q_M) \in \mathbb{Z}\left[ q_1^{\pm}, q_2^{\pm},\cdots ,q_M^{\pm}\right],
\]
which satisfies $f(p_1,\cdots,p_M)=g(q_1\cdots, q_M)$.
In these settings, the function $g$ is decomposed as
\[
g(q_1,\cdots, q_M)=p_1^{r_1}p_2^{r_2}\cdots p_M^{r_M}\cdot \tilde{g}(q_1,\cdots, q_M),
\]
where $r_1,r_2, \cdots, r_M\in\mathbb{Z}$ and $\tilde{g}(q_1,\cdots,q_M)$ is irreducible in $\mathbb{Z} \left[ q_1^{\pm}, q_2^{\pm},\cdots ,q_M^{\pm}\right]$.
\end{Lemma}
The underlying idea is the fact in algebra that the localization of a unique factorization domain preserves the irreducibility of its elements.
Proof is found in reference \cite{dKdVSC2}.

\end{document}